\documentclass[published]{JHEP3} 

\JHEP{00(2004)000}

\usepackage{epsfig,multicol}

\newcommand\fverb{\setbox\pippobox=\hbox\bgroup\verb}
\newcommand\fverbdo{\egroup\medskip\noindent%
			\fbox{\unhbox\pippobox}\ }
\newcommand{\be}{\begin{equation}}
\newcommand{\ee}{\end{equation}}
\newcommand\fverbit{\egroup\item[\fbox{\unhbox\pippobox}]}
\newbox\pippobox

\title{LMA and sterile neutrinos: a case for resonance spin flavour
precession?}

\author{Bhag C. Chauhan \thanks{On leave from Govt. Degree College, 
Karsog (H P) India 171304.}
		and Jo\~{a}o Pulido\\
   Centro de F\'\i sica das Interac\c c\~oes Fundamentais (CFIF) \\
 Departamento de F\'\i sica, Instituto Superior T\'ecnico \\
Av. Rovisco Pais, P-1049-001 Lisboa, Portugal\\
	E-mail: \email{chauhan@cfif.ist.utl.pt}, \email{pulido@cfif.ist.utl.pt
}}
\received{...} 		
\revised{...}
\accepted{...}		

\preprint{\hepph{0402194}}	

\abstract{Open questions remain after the confirmation of LMA as 
the dominant solution
to the solar neutrino deficit. These are the apparent time modulation of the
solar neutrino event rates in the Homestake, Gallium and SuperKamiokande
experiments, possibly related to solar magnetic activity, the 
discrepancy between the event rate in the Homestake experiment and its
theoretical prediction and the absence of the electron spectrum upturn
in SuperKamiokande at energies below 6-8 MeV. We search for a possible 
understanding of these questions in the context of resonant spin flavour
precession to sterile neutrinos, assuming a class of magnetic field profiles
anchored in the upper radiation/lower convection zone. We consider the
simplest such model beyond the standard 2$\nu$ flavour LMA one, with one 
single magnetic moment transition between active and sterile state and
vanishing vacuum mixing. The preferred mass square difference is
$\Delta m^2_{10}=O(10^{-8}eV^2)$. The Ga rate appears to be the most 
sensitive of all to solar activity. It is also found that a field profile 
extending within a longer region in the radial direction is favoured over 
another with a shorter span, and leads to a stronger suppression than in 
the LMA case of the intermediate 
energy neutrinos and some of the $^8 B$ ones.}

\begin{document} 


\section{Introduction}

The KamLAND experiment \cite{Eguchi:2002dm} seems to confirm the Large
Mixing Angle (LMA) solution to the solar neutrino deficit, originally presented 
in 1992 \cite{LMA}, and to rule out spin flavour precession (SFP) \cite{SFP}. 
Even the possibility of a sub-dominant SFP is substantially 
constrained \cite{Miranda} in view of the recent upper
bound on solar antineutrinos corresponding to 0.028\% of the $^8 B$ neutrino 
flux obtained by the KamLAND collaboration \cite{Eguchi:2003}. It would be 
however premature to assert that LMA \cite{LMA1} is the final and complete 
solution, at a time when solar neutrino experiments enter a new era of precision
measurements and there are hints of possible phenomena which fail to
fit in this scenario. 

One of the generic predictions of LMA is an exceedingly large rate \cite{LMA2} for 
the Cl experiment \cite{Cl}. This discrepancy reaches an excess of 2.5 $\sigma$ if
one uses the latest BP'04 results \cite{Bahc}.  
Moreover, despite accurate predictions for the SuperKamiokande 
\cite{SK} and SNO \cite{SNO} total reduced rates, the corresponding LMA spectrum
predictions exhibit an upturn below (6-8) MeV which is absent in both data.
While the absolute values of the rates may not be significant, because they
involve a normalization to a specific $^8 B$ flux whose value is known within a 
(15-20)\% theoretical uncertainty, the absence of this upturn in the data 
may motivate new physics beyond LMA, since it does not rely on normalization to a 
specific model\footnote{The present statistics both in SK and SNO are however not 
enough to perform a strict statement concerning acceptance or rejection of a model 
only on grounds of this upturn.}.   
Reconciling the theory with the data in this aspect implies a decrease in the
survival probability for neutrinos with energies in the range (0.8-8) MeV, that 
is, those which are observed by Chlorine and unobserved by SK and SNO. 

Another hint which may point to new physics lies in the possible 
time dependence of the neutrino signal. Solar neutrino statistical analyses 
confronting theoretical model predictions with data have been using time averaged 
event rates or fluxes \cite{LMA1,CP} and could, for this reason, be 
missing important information. The examination of the data on a time
basis by the Stanford Group has in fact provided increasing evidence 
that the solar neutrino flux is not constant, but varies with well-known solar 
rotation periods \cite{SS,SW,CS1}. Such a situation, if confirmed, can in no way
be explained by the LMA scenario. It can neither be understood in terms of an 
SFP transition of $\nu_e$ into $\bar\nu_{\mu}$ or $\bar\nu_{\tau}$ \cite{Miranda},
as this would originate a sizable and unseen $\bar\nu_{e}$ flux.  

The intermediate energy dip in the survival probability and the possibility of
modulation of the neutrino signal constitute our main motivation to investigate 
scenarios with a sterile neutrino. A magnetic moment driven conversion from active
to sterile neutrinos and the fact that these leave no trace in the
detectors appears to be an attractive possibility to generate these two effects.
In fact, time dependence requires this conversion to be
related to solar magnetic activity or to solar rotation with an axially 
asymmetric magnetic field profile. 

Our purpose in this paper is to perform a combined prediction of all solar
neutrino data, namely the rates for Chlorine \cite{Cl} and Gallium \cite{Ga}, 
\cite{SAGE} experiments, the SuperKamiokande \cite{SK} reduced rate and spectrum
and the reduced rates in SNO data \cite{SNO}. We will not
be concerned at this preliminary stage with $\chi^2$ fittings, but rather 
with showing the change in the LMA probability shape and solar neutrino
predictions resulting from a possible time dependent magnetic field
related conversion to the sterile neutrino. In section 2 we present the
general structure of the model, in section 3 we derive the numerical predictions
of all rates and discuss other consequences of the model for two specific solar 
field profiles and in section 4 we draw our main conclusions.  

\section{Structure of the Model}

In order to expound our model, we consider a system of two active neutrinos
and a sterile one which mix in the mass eigenstates $\nu_0$, $\nu_1$ and $\nu_2$.  
This can be parametrized by the following rotation matrix \cite{HS1}
\be
\left(\begin{array}{c}\nu_{s}\\ \nu_{e}\\ \nu_{x}\end{array}\right)=
\left(\begin{array}{ccc}c_{\alpha}&s_{\alpha}&0\\ -s_{\alpha}c_{\theta}&
c_{\alpha}c_{\theta}&s_{\theta}\\ s_{\alpha}s_{\theta}&-c_{\alpha}s_{\theta}&
c_{\theta}\end{array}\right)\left(\begin{array}{c}\nu_{0}\\ \nu_{1}\\
\nu_{2}\end{array}\right) 
\ee
with $\theta$ denoting the usual LMA vacuum mixing angle and $\alpha$ the 
vacuum sterile mixing. A straightforward but tedious calculation then leads to
the following form of the vacuum Hamiltonian
\be
\cal{H}_{\rm {vac}}=\left(\begin{array}{ccc}\frac{-\Delta m^2_{10}}{2E}c^2_{\alpha}&
\frac{\Delta m^2_{10}}{4E}s_{2\alpha}c_{\theta}&\frac{-\Delta m^2_{10}}{4E}
s_{2\alpha}s_{\theta}\\ \frac{\Delta m^2_{10}}{4E}s_{2\alpha}c_{\theta}&~
\frac{\Delta m^2_{21}}{2E}s^2_{\theta}-\frac{\Delta m^2_{10}}{2E}s^2_{\alpha}c^2_{\theta}~&
~(\frac{\Delta m^2_{21}}{4E}+\frac{\Delta m^2_{10}}{4E}s^2_{\alpha})s_{2\theta}\\
\frac{-\Delta m^2_{10}}{4E}s_{2\alpha}s_{\theta}&~(\frac{\Delta m^2_{21}}{4E}+
\frac{\Delta m^2_{10}}{4E}s^2_{\alpha})s_{2\theta}~&~\frac{\Delta m^2_{21}}{2E}c^2_{\theta}-
\frac{\Delta m^2_{10}}{2E}s^2_{\alpha}s^2_{\theta}\end{array}\right)  
\ee   
with obvious notation. We seek for a model which should be the simplest departure 
from the conventional LMA case and be able to generate a time dependent transition
into a sterile neutrino. To this end we take a vanishing mixing angle, $\alpha$=0,
so that active states $\nu_e,~\nu_{\mu}$ communicate to the sterile one through
magnetic moment transitions only. The matter Hamiltonian is thus
\be
\cal{H}_{\rm {M}}=\left(\begin{array}{ccc}\frac{-\Delta m^2_{10}}{2E}&
\mu_{1}B&\mu_{2}B\\ \mu_{1}B& \frac{\Delta m^2_{21}}{2E}s^2_{\theta}+V_e&
\frac{\Delta m^2_{21}}{4E}s_{2\theta}\\ \mu_2{B}&\frac{\Delta m^2_{21}}{4E}s_{2\theta}&
\frac{\Delta m^2_{21}}{2E}c^2_{\theta}+V_{\mu}\end{array}\right)
\ee
with $V_e$ and $V_{\mu}$ being the matter induced potentials for $\nu_e$ and $\nu_{\mu}$
respectively and $\mu_{1}$, $\mu_{2}$ their transition moments to the sterile 
neutrino. We will consider field profiles concentrated around the bottom of the convective 
zone, as motivated by the dynamo theory \cite{SS,SW,CS1}, and hence require spin flavour 
precession to be resonant in this region. 
Therefore the two processes (LMA and RSFP) occur sequentially at very different solar radii,
with the RSFP critical density determined by a mass square difference between one
of the active states and the sterile one $O(10^{-8}eV^2)$. An efficient conversion to 
the sterile neutrino requires that the transition moment associated with this order of
magnitude difference dominates over the other.  
Given these facts we choose $\Delta m^2_{10}=O(10^{-8}eV^2)$
and so $\mu_{2}=0$. \footnote{The other situation, namely a non-vanishing $\mu_{2}$ and
$\mu_{1}=0$ would correspond very closely to the LMA one and two sizable transition 
moments to an intermediate possibility.}   

We obtain the neutrino survival probability ($P(\nu_e \rightarrow \nu_e)$=$P_{ee}$)
by assuming the propagation inside the sun to be adiabatic except in the region 
where the adiabaticity parameter reaches its minimum. This is essentially a 
generalization of the (2x2) case proposed some time ago \cite{Parke}. In fig.1 we
show the evolution of the mass matter eigenvalues inside the sun for a typical case. 
It corresponds to a neutrino energy $E=2.1MeV$, $\Delta m^2_{10}=-1.7\times10^{-8}eV^2$ 
and field profile 2 described below with peak field $B_{0}=3.0\times 10^{5}~G$. Of 
the three eigenvalues, it is seen that $\nu_1$ and 
$\nu_0$ become the closest in the convection zone ($r/R_S\simeq 0.8$), so that adiabaticity
may be broken in the transition at this critical density. 
We recall that a strong field in 
this region will provide an efficient, adiabatic conversion corresponding to a larger 
eigenvalue separation while a weaker field will generate less conversion, with closer 
eigenvalues. Moreover the LMA oscillation and the $\nu_2 \rightarrow \nu_0$ transition 
are strongly adiabatic. This suggests that a field profile exhibiting a time dependent 
peak value deep inside the convective zone will manifest itself in the modulation of 
the neutrino signal. Hence we need only to consider the jump probability $P_C$ between 
$\nu_1$ and $\nu_0$ and the expression describing the three neutrino propagation is
\be
\left(\begin{array}{c}\nu_{s}(t)\\ \nu_{e}(t)\\ \nu_{\mu}(t) 
\end{array}\right)\!\!=~~
[R_{o}^{T}]~~[A]~~[B]~~[C]~~
[R_{i}]~~\left(\begin{array}{c}0 \\ 1\\ 0 
\end{array}\right)
\ee
where
\be
[A]=\left(\begin{array}{ccc}\!\!e^{-i\int_{r_{_x}}^{r_o}\!\! H_{D_{11}}dr}\!\!& & \\ & \!\!
e^{-i\int_{r_{_x}}^{r_o}\!\! H_{D_{22}}dr}\!\!& \\ & &\!\!
e^{-i\int_{r_{_x}}^{r_o}\!\! H_{D_{33}}dr}\!\!\end{array}\right)
\ee
\be
[B]=\left(\begin{array}{ccc}\sqrt{1-P_C}&\sqrt{P_C}&0\\ \sqrt{P_C}&\sqrt{1-P_C}&0\\
0&0&1 \end{array}\right)
\ee

\begin{figure}[h]
\setlength{\unitlength}{1cm}
\begin{center}
\hspace*{-1.6cm}
\epsfig{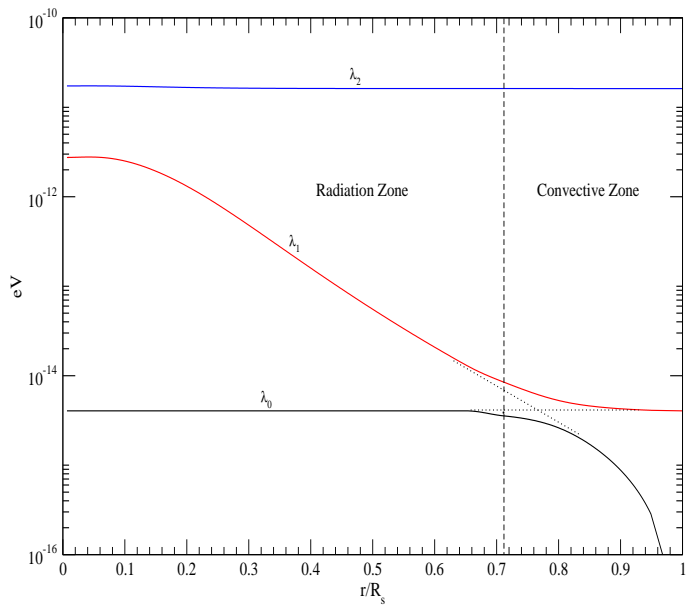}
\end{center}
\caption{ {\bf The evolution of the mass matter eigenvalues along the solar 
neutrino trajectory.} The example chosen is for profile 2 (see fig.2) 
with neutrino energy $E=2.1MeV$,
$\Delta m^2_{10}=-1.7\times 10^{-8}eV^2$, $B_0=3.0\times 10^5G$.
LMA parameters are slightly shifted from their best fit values \cite{SNO}:
$\Delta m^2_{21}=6.8\times 10^{-5}eV^2$, $\theta=32.1^{o}$.} 
\label{fig1}
\end{figure}

\be
[C]=\left(\begin{array}{ccc}\!\!e^{-i\int_{r_i}^{r_{_x}}\!\! H_{D_{11}}dr}\!\!& & \\ & \!\!
e^{-i\int_{r_i}^{r_{_x}}\!\! H_{D_{22}}dr}\!\!& \\ & &\!\!
e^{-i\int_{r_i}^{r_{_x}}\!\! H_{D_{33}}dr}\!\!\end{array}\right)
\ee
Here $H_{D}$ denotes the diagonalized Hamiltonian, $r_i$ the neutrino production point, $r_o$
the detection point and $r_{_x}$ the location of the RSFP critical density. $[R]$ is the
rotation matrix from the weak eigenstate to the matter eigenstate basis. Its form in
the vacuum is given by the transpose of (2.1) with $\alpha=0$. In matter with magnetic
field it becomes in its most general form
\be
[R]=\left(\begin{array}{ccc}c_{01}c_{02}&~~-\!s_{01}c_{12}\!-\!c_{01}s_{12}s_{02}~~&
~~s_{01}s_{12}\!-\!c_{01}c_{12}s_{02}\\ s_{01}c_{02}&~~c_{01}c_{12}\!-\!s_{01}s_{12}s_{02}
~~&~~-\!c_{01}s_{12}\!-\!s_{01}c_{12}s_{02}\\ s_{02}&s_{12}c_{02}&c_{12}c_{02}
\end{array}\right)
\ee
where $\theta_{01}$, $\theta_{02}$ are the mixing angles induced by the neutrino magnetic 
moments $\mu_1$, $\mu_2$ respectively. The matter mixing angle $\theta_{12}$ becomes
in the vacuum the LMA mixing angle $\theta$, eq. (2.1).
In equation (2.4) $[R_i]$ converts the weak eigenstate $\nu_e$ produced in
the solar core into a combination of Hamiltonian eigenstates which propagate adiabatically
(matrices [C] and [A]) except for the correction at the critical density described by
the matrix [B] where $\nu_1$ and $\nu_0$ nearly meet. In order to account for level 
crossing at the resonance between states $\nu_0,~\nu_1$, the corresponding first two
rows of the rotation matrix $R$ are interchanged so as to obtain $R_{o}$.   
Hamiltonian eigenstates are then converted back to the weak basis through $R_{o}^{T}$ 
at the detection point. Taking $\theta_{02}=0$ ($\mu_2=0$) and using eq.(2.8), matrices 
$[R_i]$ and $[R_{o}^{T}]$ therefore read 
\be
[R_i]=\left(\begin{array}{ccc}c_{01}&-s_{01}c_{12}&s_{01}s_{12}\\
s_{01}&c_{01}c_{12}&-c_{01}s_{12}\\0&s_{12}&c_{12}\end{array}\right)
\ee
\be
[R_{o}^{T}]=\left(\begin{array}{ccc}s_{01}&c_{01}&0\\c_{01}c_{12}&-\!s_{01}\!c_{12}&~s_{12}\\
-c_{01}s_{12}&s_{01}\!s_{12}&c_{12}\end{array}\right)
\ee

For the jump probability $P_C$ we use the Landau Zener approximation
\be
P_C=exp(-\frac{\pi}{2} \gamma_{_C})
\ee
with the adiabaticity parameter given by the ratio between the corresponding 
eigenvalue difference and the spatial rate of the mixing angle 
\be
\gamma_{_C}=\left |\frac{\lambda_0-\lambda_1}
{2\dot\theta_{01}}\right|.
\ee
Using equations (2.4) - (2.10) and the fact that at the neutrino production
and detection points the absence of magnetic field implies the vanishing of angle
${\theta_{01}}$, the expressions for the survival and
conversion probabilities become
\be 
P_{ee}=|(\nu_e(t),\nu_{e})|^2=c_{\theta}^2{c_{12}^{2}}_{i}P_C+s_{\theta}^2{s_{12}^{2}}_{i}  
\ee
\be
P_{e\mu}=|(\nu_{\mu}(t),\nu_{e})|^2=s_{\theta}^2{c_{12}^{2}}_{i}P_C+c_{\theta}^2{s_{12}^{2}}_{i}
\ee
where we neglected fast oscillating terms which average to zero and subscript $i$
refers to the mixing angle at the production point.
The LMA survival and conversion probabilities in the 2-flavour case
can be readily obtained from eqs. (2.13), (2.14) for 
$P_C=1$ (vanishing magnetic field at the resonance point). 
  
The expression for the SK spectrum to be used is 
\be
R^{th}_{SK}=\frac{\int_{{E_e}_{min}}^{{E_e}_{max}}dE_e\int_{m_e}^{{E^{'}_e}_{max}}dE^{'}\!\!_e
f(E{'}_e,E_e)\int_{E_m}^{E_M}dE\phi(E)[P_{ee}(E)\frac{d\sigma_{\nu_e}}{dT^{'}}+P_{e\mu}(E)
\frac{d\sigma_{\nu_{\mu}}}{dT{'}}]}
{\int_{{E_e}_{min}}^{{E_e}_{max}}dE_e\int_{m_e}^{{E^{'}_e}_{max}}dE^{'}\!\!_e
f(E{'}_e,E_e)\int_{E_m}^{E_M}dE\phi(E)\frac{d\sigma_{\nu_e}}{dT^{'}}}
\ee
with standard notation \cite{Astrop} and a similar expression for the SNO one. We
take the SK and SNO energy resolution functions from \cite{Fukuda1998},\cite{SNOhom}.

\newpage

\begin{figure}[h]
\setlength{\unitlength}{1cm}
\begin{center}
\hspace*{-1.6cm}
\epsfig{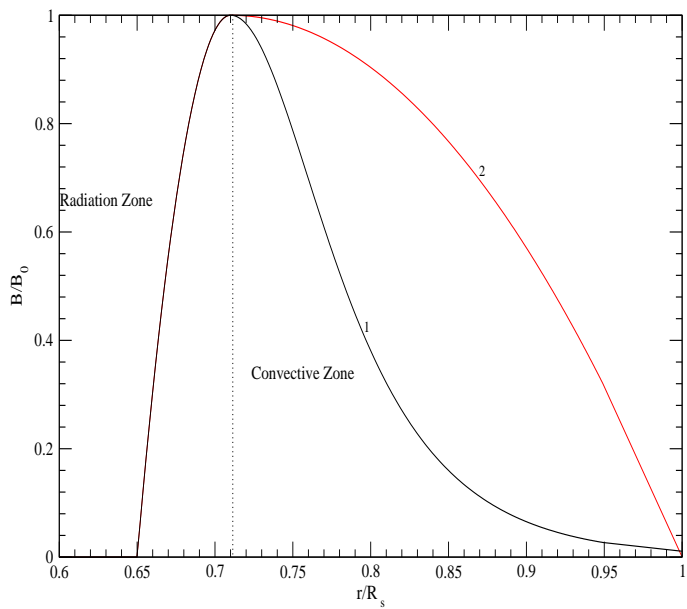}
\end{center}
\caption{ {\bf Solar magnetic field profiles}. These are motivated by the 
dynamo process near the tachocline \cite{CS1}: they exhibit a common and
fast rise starting at the upper radiation zone reaching a peak at the
bottom of the convection zone. Profile 2 extends over a longer region,
hence it leads to more efficient sterile neutrino conversion.}
\label{fig2}
\end{figure}

\section{Solutions and Discussion} 

The two classes of field profiles to be used are characterized by a sharp increase
starting in the radiative zone just below the tachocline,
reaching a peak at the bottom of the convective zone and a
smoother decrease up to the solar surface. They are (see fig.2) 

$Profile~1$
\be
B=B_0\left(1-\left(\frac{x-x_{_C}}{0.06}\right)^2\right)~~,~~x_{_R}\leq x \leq x_{_C}
\ee
\be 
B=\frac{B_0}{cosh(18(x-x_{_C}))}~~,~~x_{_C}<x \leq 1
\ee

$Profile~2$
\be
B=B_0\left(1-\left(\frac{x-x_{_C}}{0.06}\right)^2\right)~~,~~x_{_R}\leq x \leq x_{_C}
\ee


\be
B=B_0\left(1-\left(\frac{x-x_{_C}}{1-x_{_C}}\right)^2\right)~~,~~x_{_C}<x \leq 1
\ee
with vanishing field for $x<x_{_R}$. Here $x$ is the fraction of the solar radius, 
$x_{_R}=0.65$, $x_{_C}=0.71$ and we take in all cases a magnetic moment 
$\mu_1=10^{-12}\mu_{B}$. The preferred peak field values are found to be in the range 
$B_0=(2-3)\times10^5G$.

\begin{figure}[h]
\setlength{\unitlength}{1cm}
\begin{center}
\hspace*{-1.6cm}
\epsfig{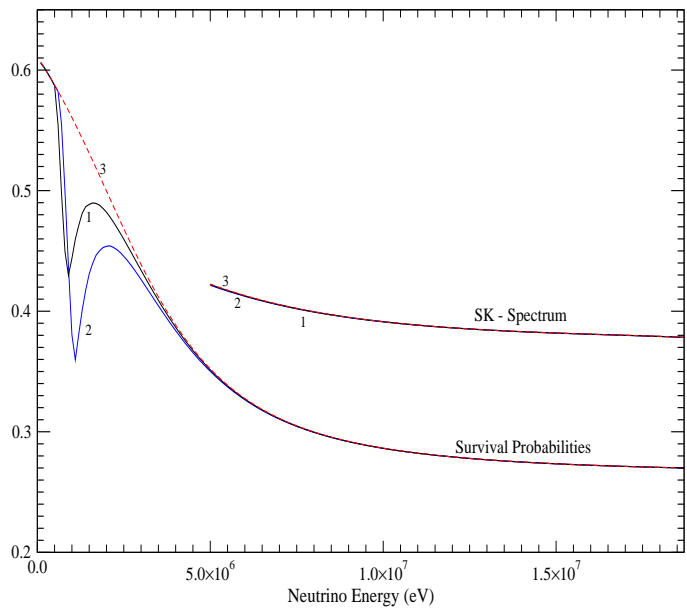}
\end{center}
\caption{ {\bf Survival probabilities and SuperKamiokande spectrum for profile 1 
(curves 1 and 2).} For comparison, the LMA curves are also given
(curves 3). The dip in the probability hardly affects the shape of the 
spectrum. (See also table II).} 
\label{fig3}
\end{figure}

In table I we show the solar neutrino data on rates and their comparison
with the SSM \cite{Bahc} predictions including the theoretical uncertainties \footnote{
Here we use the BP04 standard solar model \cite{Bahc}.}. We refer to the parameter values
as given from global fits \cite{SNO} $\Delta m^2_{21}=7.1\pm^{1.2}_{0.6}\times 10^{-6}eV^2$,
$\theta =32.5\pm^{2.4}_{2.3}$ degrees. Our main results are summarized in figs.3 - 6 and 
tables II, III. 

\begin{center}
\begin{tabular}{lcccc} \\ \hline \hline
Experiment &  Data      &   Theory   &   Data/Theory  &  Reference \\ \hline
Homestake  &  $2.56\pm0.16\pm0.15$ & $8.5\pm^{1.8}_{1.8}$ & $0.301\pm{0.069}$ &
\cite{Cl} \\
SAGE     &  $70.9\pm^{5.3}_{5.2}\pm^{3.7}_{3.2}$ & $131\pm ^{12}_{10}$ & $0.541
\pm^{0.070}_{0.062}$ & \cite{SAGE} \\
Gallex+GNO & $70.8\pm{4.5}\pm{3.8}$ & $131\pm ^{12}_{10}$ & $0.540
\pm^{0.067}_{0.061}$ & \cite{Ga}\\
SuperKamiokande&$2.35\pm{0.02}\pm{0.08}$ &
$5.82\pm{1.34}$&$0.404\pm{0.094}$& 
\cite{SK}\\
SNO CC &$1.59\pm^{0.08}_{0.07}\pm^{0.06}_{0.08}$&$5.82\pm{1.34}$&
$0.273\pm{0.065}$& \cite{SNO} \\ 
SNO ES &$2.21\pm^{0.31}_{0.26}\pm{0.10}$&$5.82\pm{1.34}$&
$0.380\pm^{0.104}_{0.100}$& \cite{SNO}\\
SNO NC & $5.21\pm{0.27}\pm{0.38}$&$5.82\pm{1.34}$&$0.895\pm{0.221}$
& \cite{SNO}\\ \hline
\end{tabular}
\end{center}

{{\bf Table I - Data from the solar neutrino experiments.} Units are SNU for
Homestake and Gallium and $10^{6}cm^{-2}s^{-1}$ for SuperKamiokande and SNO. 
The uncertainties in the fourth column include solar standard model errors.} 
\begin{figure}[h]
\setlength{\unitlength}{1cm}
\begin{center}
\hspace*{-1.6cm}
\epsfig{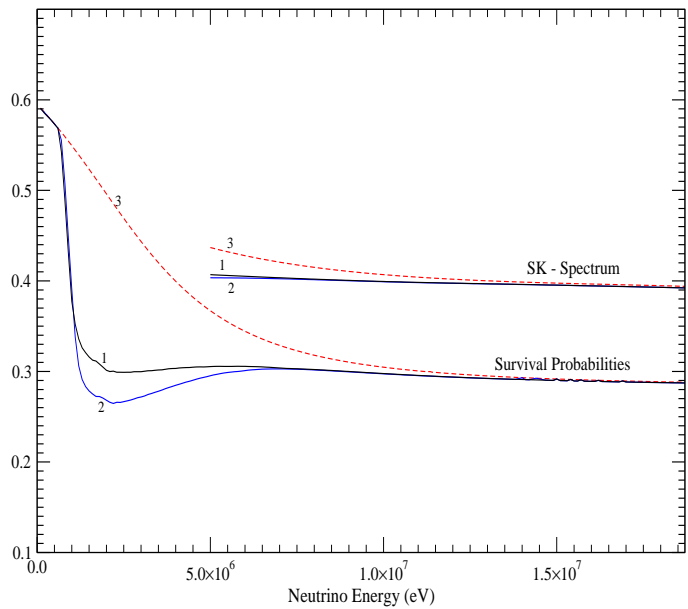}
\end{center}
\caption{ {\bf Same as fig.3 for profile 2.} A downturn of the spectrum appears
in this case as some $^8 B$ neutrinos ($E>5MeV$) are converted to sterile ones. 
(See also table III).  }
\label{fig4}
\end{figure}

\begin{center}
\begin{tabular}{cccccccc} \\ \hline \hline
$\Delta m^2_{10}~(eV^{2})$&$B_0(G)$&Cl&Ga&SK&$SNO_{CC}$&$SNO_{ES}$&$SNO_{NC}$ \\ \hline 
$-1.35\times10^{-8}eV^{2}$&$2.0\times10^{5}G$&2.77&66.5&0.406&0.284&0.481&0.999 \\
$-1.65\times10^{-8}eV^{2}$&$2.6\times10^{5}G$&2.76&66.9&0.406&0.284&0.476&0.999 \\ \hline
\end{tabular}
\end{center}
{{\bf Table II - Profile 1: Model predictions for rates.} Units are SNU for Cl, Ga 
and the ratio of the model prediction by the SSM value for SK and SNO. See also fig.3 
for the survival probabilities and SK spectrum. Here $\Delta m^2_{21}~(eV^{2})=
\vspace{0.5cm}
6.5\times 10^{-5}eV^2$, $\theta=30.9^{o}$.}
\indent Fig.3 (profile 1) and fig.4 (profile 2) show the electron neutrino survival probability
and the electron energy spectrum in SK for two values of $\Delta m^2_{10}$ in each
profile together with the LMA probability and spectrum,
for comparison, near the pure LMA best fit point.
It is seen that the upturn in the LMA spectrum can be reduced or made to disappear.
A flat spectrum, as suggested by the data \cite{SK}, is in fact obtained for a 
convenient choice of $\Delta m^2_{10}$ and $B_0$ in the case of profile 2 but not for  
profile 1. For both profiles, however, the Cl rate is substantially decreased from its 
LMA prediction of approximately 3.1 SNU (see tables II, III). Parameter 
$\Delta m^2_{10}$ also determines the 
location of the important $\nu_{1} \rightarrow \nu_{0}$ 
resonance, with less negative 
values corresponding to resonances closer to the solar surface. 
Also the longer spatial 
extension of a strong field in the case of profile 2 implies more conversion efficiency, 
a fact which is clearly reflected in a wider dip in the probability curve and even a 
flatness in the spectrum for $\Delta m^2_{10}$=$-(1.6-1.7)\times10^{-8}~eV^2$. All
rates can be brought to within $1\sigma$ of the experimental value with 
theoretical errors included. 
In view of our motivation to account for possible modulations in the rates,
we will not attempt at 
quantitative and detailed fits at this stage, since the 
solar neutrino data used are averages and further, they do not refer to the same 
periods in different experiments.   
\begin{center}
\begin{tabular}{cccccccc} \\ \hline \hline
$\Delta m^2_{10}~(eV^{2})$&$B_0(G)$&Cl&Ga&SK&$SNO_{CC}$&$SNO_{ES}$&$SNO_{NC}$ \\ \hline 
$-1.6\times10^{-8}eV^{2}$&$2.6\times 10^{5}$&2.76&65.1&0.403&0.297&0.431&0.982 \\
$-1.7\times10^{-8}eV^{2}$&$3\times10^{5}$&2.76&65.5&0.402&0.297&0.421&0.981 \\ \hline
\end{tabular}
\end{center}
{{\bf Table III - Same as table II for profile 2.} See also fig.4 for the survival 
probabilities and SK spectrum. Here $\Delta m^2_{21}~
\vspace{0.5cm}
(eV^{2})=6.8\times10^{-5}eV^2$, $\theta=32.1^{o}$.}

We note that the predictions from profiles 1 and 2 exhibit a manifest difference: 
in fact for profile 2 increasing the dip in the probability through an increase
in the peak field leads to a stronger downturn of the spectrum, while for profile 1 
an increase in the peak field leaves the spectrum unaffected (figs.3 and 4).
This is because in profile 2 a higher peak field leads to a stronger suppression of 
the low energy sector of the $^8 B$ neutrinos, 
which are present in the SK rate, along 
with pep and CNO ones. For profile 1, which is more localized, an increase in the 
peak field affects only the suppression of the 
intermediate energy neutrinos in Cl, especially the pep neutrinos with energy 
$E_{pep}=1.44MeV$ and the CNO ones, all absent in the SK rate and accounting for only 
a minor fraction of the Cl rate. 
\begin{figure}[h]
\begin{center}
\hspace*{-2.0cm}
\epsfig{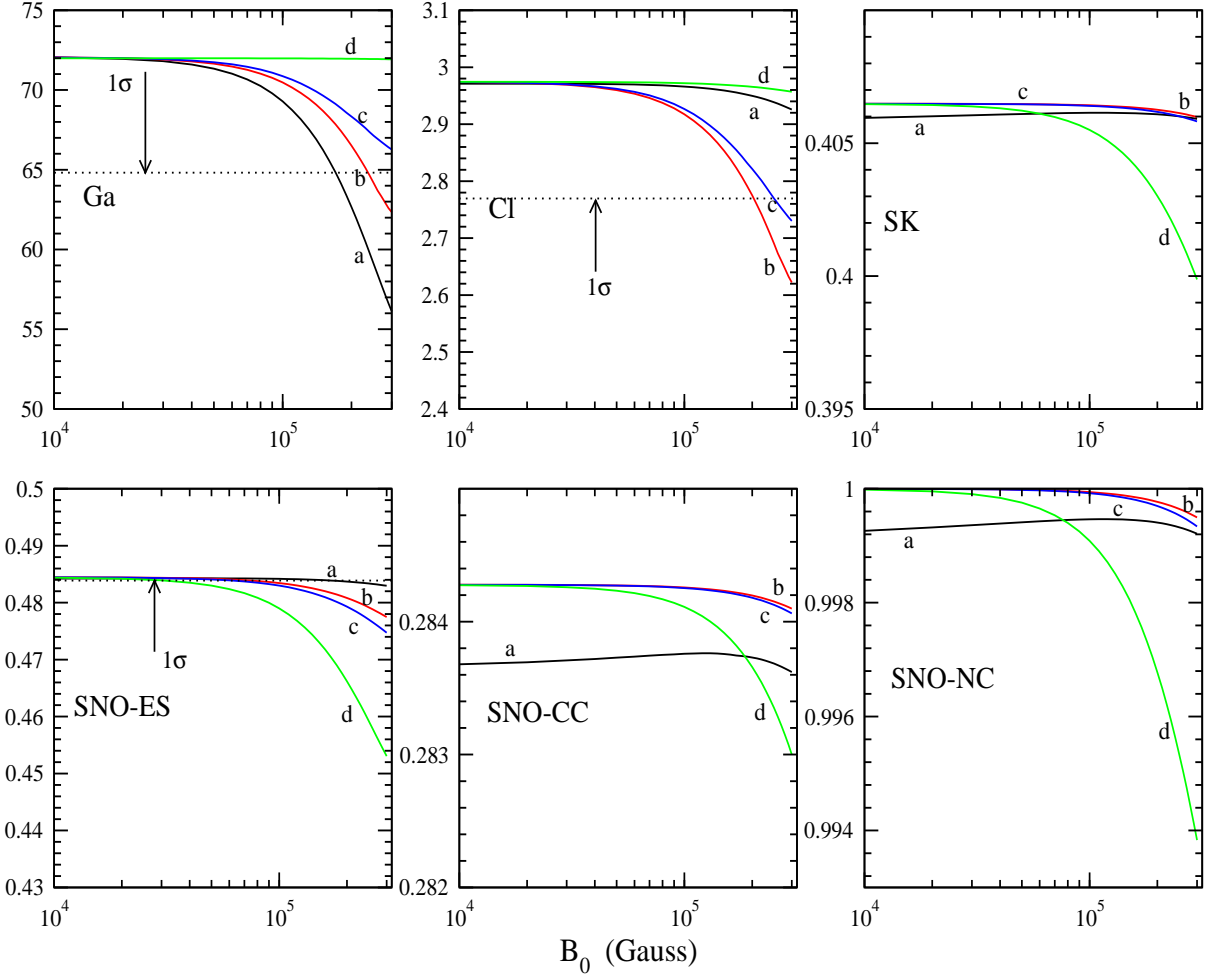}
\end{center}
\caption{ {\bf Event rates as a function of the peak field $B_0$ for profile 1.}   
Variations of solar activity, manifested in a variable field intensity, are
reflected in variations of the event rates. SK, SNO-CC and SNO-NC panels are
all within $1\sigma$ of the expeimental value including theoretical errors. The 
LMA parameters are slightly shifted from their best fit values \cite{SNO}. (See
also table IV).}
\label{fig5}
\end{figure}

In figs.5, 6 we show the dependence of rates on the peak field value for profiles
1, 2, from a starting point with unaffected LMA predictions, up to $B_0=3\times10^5G$
\cite{Antia} with a view on a possible anticipation of future studies of time
dependence in neutrino signals. Four different cases of $\Delta m^2_{10}$ in the
relevant range $(10^{-8}eV^2)$ are considered (see table IV).
The values of $\Delta m^2_{21}$ and $\theta$ are slightly different in figs.5 and 6, 
hence the small discrepancy in the rates in the pure LMA limit. They are however well 
within their $1\sigma$ range \cite{SNO} (see the captions of tables II, III). We find 
that the Ga rate is the most sensitive of all to 
\begin{figure}[h]
\begin{center}
\hspace*{-1.6cm}
\epsfig{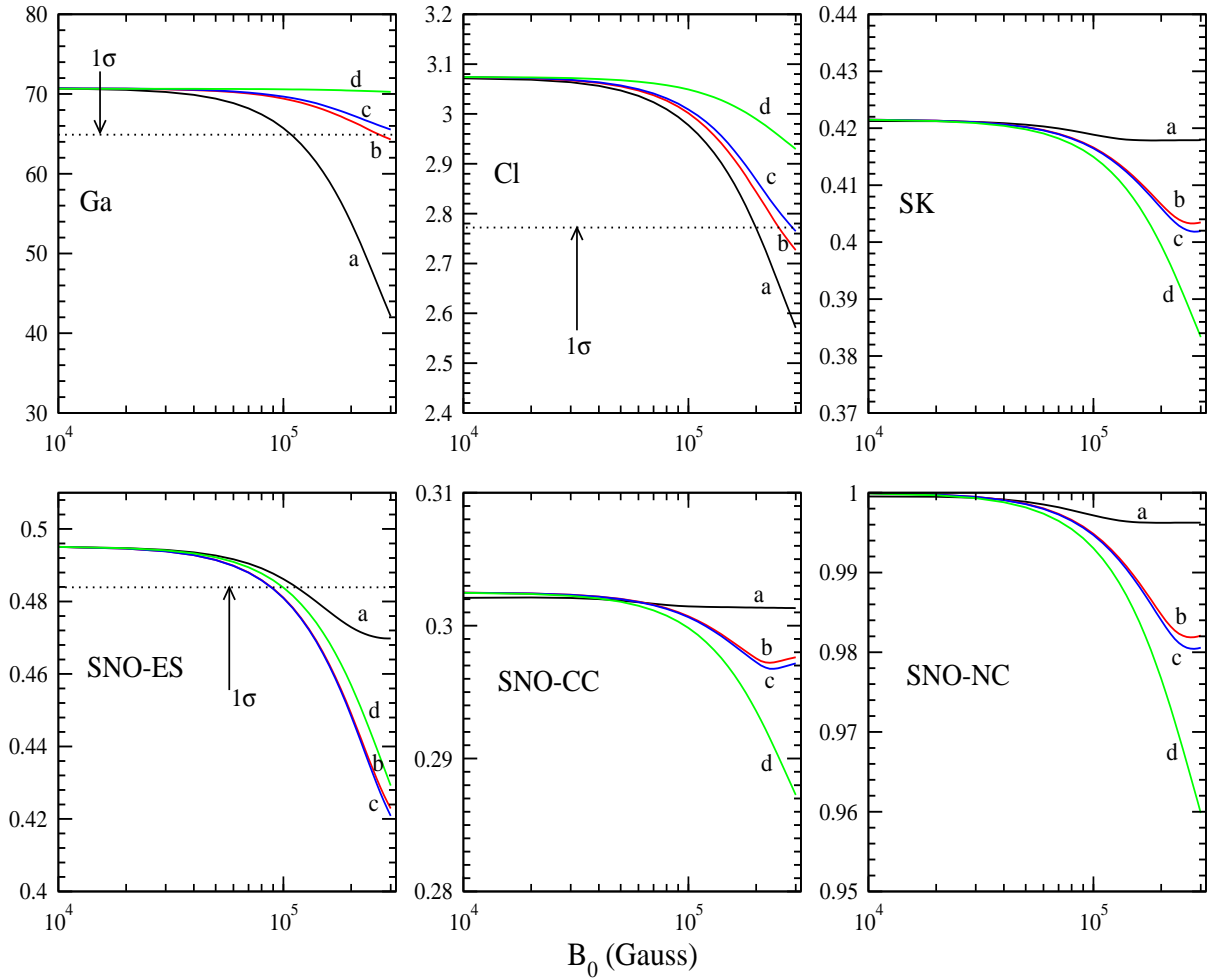}
\end{center}
\caption{ {\bf Same as fig.5 for profile 2.} The LMA parameters are slightly 
shifted from their best fit values \cite{SNO}. (See also table IV).}
\label{fig6}
\end{figure}
changes in $B_0$ with a maximum 
modulation for profiles 1 and 2 respectively
\be
\frac{\Delta R_{Ga}}{R_{Ga}}\simeq 22\%~,~\frac{\Delta R_{Ga}}{R_{Ga}}\simeq 40\%
\ee
At this point it should be noted that
in their detailed analysis, the Stanford Group \cite{SS} finds the Ga rate to
exhibit the most manifest time dependence of all with two clear peaks in the ranges 
45-75 and 90-120 SNU. It is unclear at this stage whether this apparent convergence
of results is essential or not, a situation which obviously deserves further 
investigation.
\begin{center}
\begin{tabular}{ccccc} \\ \hline \hline
&a&b&c&d \\ \hline
Profile 1&$-0.5\times 10^{-8}eV^2$&$-1.35\times 10^{-8}eV^2$&$-1.65\times 10^{-8}eV^2$&
$-5\times 10^{-8}eV^2$ \\ \hline
Profile 2&$-0.5\times 10^{-8}eV^2$&$-1.6\times 10^{-8}eV^2$&$-1.7\times 10^{-8}eV^2$&
$-5\times 10^{-8}eV^2$ \\ \hline
\end{tabular}
\end{center}
{{\bf Table IV - The values of $\Delta m^2_{10}$ labeled a, b, c, d in figures 5 and 6.}}
\\
\newpage
\indent The Cl rate is less sensitive with a maximum
\be
\frac{\Delta R_{Cl}}{R_{Cl}}\simeq 12\%~,~\frac{\Delta R_{Cl}}{R_{Cl}}\simeq 17\%
\ee
for profile 1 and profile 2 respectively. For all other rates this sensitivity
is always below 7\% for profile 1 and 15\% for profile 2. All these modulations
depend of course on the value of $\Delta m^2_{10}$ chosen. Furthermore the relative
sensitivities of different rates to $\Delta m^2_{10}$ change from one rate to another.  
The cases described in detail in tables II, III and figs.3, 4 can also be seen in 
a different context in figs. 5, 6. 

In all our calculations we have used a neutrino magnetic moment $\mu_{\nu}=10^{-12}
\mu_B$. This is larger than the supernova bound \cite{SN,SN1} by an approximate factor
of 2 which applies for active-sterile magnetic moment conversions. Hence, if 
strictly enforced, this bound would require a peak field value of $(4-6)\times 10^5G$
instead of $(2-3)\times 10^5G$. However the supernova bound is plagued by uncertainties
in the models of the supernova core \cite{SN1} or based on rough estimates 
of the supernova energetic \cite{SN}, so it cannot be expected to be quite stringent.
On the other hand the solar magnetic field profile is poorly known, both in shape and 
strength. Whereas a field of the order of $3\times10^5G$ 
is quite possible in the tachocline \cite{Antia}, it is not clear whether this can 
be exceeded by an extra factor of (1.5-2) or whether this factor could come from 
intrinsic inaccuracy of the supernova bound. 

\section{Conclusions}

To conclude, now that KamLAND has established that LMA is the dominant solution to
the solar neutrino problem, there remain a few questions, which may turn out to be 
quite important, as LMA may be incomplete and new physics might be necessary. To
this end, the hints from solar neutrinos are a possible time variation 
of the neutrino event rate \cite{SS,SW,CS1}
in the experiments, the absence of an 
electron spectrum upturn for low energies in opposition to the LMA expectation, 
and a rate in the Cl experiment lower than the LMA predicted one.   

We proposed an answer to these questions by adding a sterile state
to the two solar neutrino system. Such scenarios have already been developed in
the literature \cite{HS1,BNV} in the context of oscillations alone, in particular it
was shown that a sterile state may provide a solution to the above questions 
\cite{HS1}, except for the possible time modulation. Visible states communicate  
in those models with the sterile one via a vacuum mixing angle. In the model proposed 
in this paper, they communicate instead with the sterile one via magnetic moment 
transitions, so that time dependence of neutrino signals, if confirmed, can be 
directly connected to solar magnetic activity or to solar rotation. The new mass 
square difference is about three orders of magnitude below the LMA one, so that the 
transition to the sterile state resonates mainly in the tachocline, where the solar 
magnetic field is expected to be concentrated. In this way, a situation in which 
the transition moment associated with the smallest mass square difference dominates 
over the other leads to the most efficient conversion. 

The upturn in the spectrum can remain unchanged, be reduced or eliminated,  
depending on the choice of this mass square difference 
and the field profile. As far as the Cl rate prediction is concerned, a sizable 
reduction from its LMA value is always obtained in the class of field profiles 
investigated. Profiles 1 and 2 illustrate these situations: the elimination of the 
spectrum upturn may or may not accompany the reduction in the Cl rate (profiles 2 
and 1 respectively). It should be noted however that both SK and SNO data do not 
contain at present enough statistics that allow a strict statement on the exact 
spectrum shape.

We have also investigated the evolution of the event rates with the peak
field values in each of the two profiles and found the Ga one to be the most sensitive
of all. Interestingly enough, it was found by the Stanford Group \cite{SS} from a 
combined analysis of Gallex-GNO and SAGE that the Ga rate exhibits a strong time 
dependence with two clear peaks at 45-75 and 90-120 SNU. This apparent convergence 
between ours and their results certainly deserves further investigation. The Cl rate 
comes next to the Ga one in sensitivity, followed by all others (see figs.5, 6). In 
all cases we used the convenient mass square difference order of magnitude, 
$\Delta m^2_{10}=O(10^{-8}eV^2)$. Profile 2, with a longer spatial extension, leads 
to more sensitivity of the rates.  
We have not performed $\chi^2$ fittings, as, within the motivation of the 
present paper, they will be justified only for time dependent solar neutrino data 
which may soon become available.   

Finally, the uncertainties both in the supernova neutrino magnetic moment bound and 
in the solar field profile make it premature to discard or rule out the present 
\vspace{0.5cm}  
analysis. 
{\bf Acknowledgements} \\
{\em We acknowledge useful discussions with Peter Sturrock and David Caldwell. 
The work of BCC  was supported by Funda\c{c}\~{a}o para a 
Ci\^{e}ncia e a Tecnologia through the grant SFRH /BPD/5719/2001.}


\newpage


\end{document}